\documentstyle[preprint,aps,epsfig]{revtex} 
\tighten
\begin{document}
\small 
\preprint{\sf OUTP-96-48P}
\draft 
\title{Natural Supergravity Inflation}
\author{\large Jennifer A. Adams \vspace{2mm}}
\address{Department of Theoretical Physics, Uppsala University, \\ 
         Box 803, S-75108 Uppsala, Sweden}
\author{{\large Graham G. Ross\ \&\ Subir Sarkar}~\footnote{PPARC
 Advanced Fellow \\ \mbox{ } {\sf hep-ph/9608336}} \vspace{2mm}}
\address{Theoretical Physics, University of Oxford, \\
         1 Keble Road, Oxford OX1 3NP, U.K. \vspace{5mm}}
\date{submitted August 15, revised October 28, 1996; 
  to appear in {\em Phys. Lett. B}}
\maketitle
\begin{abstract} 
We identify a new mechanism in supergravity theories which leads to
successful inflation without any need for fine tuning. The simplest
model yields a spectrum of density fluctuations tilted away from
scale-invariance and negligible gravitational waves. We demonstrate
that this is consistent with the observed large-scale structure for a
cold dark matter dominated, critical density universe. The model can
be tested through measurements of microwave background anisotropy on
small angular scales.
\end{abstract}
\pacs{\vfill 98.80.Cq, 04.65.+e, 98.65.Dx, 98.70.Vc}
\widetext

\section{Introduction}

Field theoretic models of cosmological inflation are generically
required to provide a very flat potential for the `inflaton' field,
the large and approximately constant vacuum energy of which drives an
exponential increase of the scale-factor and is then converted to
radiation when the inflaton settles into its global minimum
\cite{inflbook}. In building such a model care must be taken to avoid
the hierarchy problem which arises because the fundamental
interactions, in particular gravity, will communicate any such high
scale of new physics to all sectors of the theory, driving the
electroweak breaking scale far above its observed value. The only
known way to avoid this problem is through the introduction of
supersymmetry which can protect the low energy scales from such
radiative corrections to all orders \cite{susyrev}. The local version
of supersymmetry --- supergravity --- incorporates gravity and has
therefore been extensively studied in attempts to construct a unified
description of all the fundamental interactions, the most ambitious of
which is the superstring.

However supergravity inflationary models \cite{susyinflrev} suffer
from their own `hierarchy' problem. The large cosmological constant
during inflation breaks supersymmetry, giving all scalar fields a soft
mass of order the Hubble parameter \cite{susybreak}. The resulting
curvature of the inflaton potential is typically too large to allow a
sufficiently long period of inflation to occur \cite{ed}. There have
been many suggested solutions to this problem but most of them are
deemed to be ad-hoc or unworkable \cite{lyth} (although inflation
driven by a D- term \cite{stewart,Dterm} may be viable). In
continuation of this discussion and our previous work \cite{success},
we wish to propose a new mechanism leading to successful inflation in
the low-energy effective supergravity theory following from the
superstring. We demonstrate that in a wide class of such theories, the
equations of motion have an infra-red fixed point at which successful
inflation can occur, even for minimal kinetic terms, along a F-flat
direction. Moreover the resulting inflationary potential has a very
specific structure, allowing precise predictions for the generated
perturbations --- both scalar and tensor. We compute these in detail
for a cold-dark matter dominated critical density universe, including
non-linear evolutionary effects, and compare with the results of the
APM galaxy survey \cite{apm}. We find reasonable agreement with the
data without having to invoke a component of hot dark matter. The
expected power spectrum of the angular anistotropy in the cosmic
microwave background (CMB) is also calculated using a Boltzmann code
and normalized to the COBE observations \cite{cobe} on large angular
scales. Ongoing and future observations on small angular scales
\cite{cmbrev} will thus provide a definitive test of the model.

\section{Requirements of the Inflationary Potential}

We begin by briefly reviewing the necessary ingredients for successful
inflation with a scalar potential $V(\phi)$. Essentially all model
generating an exponential increase of the cosmological scale-factor
$a$ satisfy the `slow-roll' conditions \cite{inflrev}
\begin{equation}
 \dot{\phi} \simeq -\frac{V'}{3H}\ , \quad
 \epsilon \equiv \frac{M^2}{2} \left(\frac{V'}{V}\right)^2 \ll 1\ ,
  \qquad
 |\eta| \equiv \left|M^2 \frac{V^{''}}{V}\right| \ll 1\ ,
\label{slowroll}
\end{equation}
where $H\simeq\sqrt{V/3M^2}$ is the Hubble parameter during inflation,
and the normalized Planck mass $M\equiv\,M_{\rm
Pl}/\sqrt{8\pi}\simeq2.44\times10^{18}$ GeV. Inflation ends
(i.e. $\ddot{a}$ drops through zero) when $\epsilon,
|\eta|\simeq1$. The spectrum of adiabatic scalar perturbations is
\cite{inflrev}
\begin{equation}
 \delta^2_{\rm H} (k) = \frac{1}{150 \pi^2} \frac{V_{\star}}{M^{4}} 
                      \frac{1}{\epsilon_{\star}}\ ,  
\label{deltah}
\end{equation}
where $\star$ denotes the epoch at which a scale of wavenumber $k$
crosses the `horizon' $H^{-1}$ (more correctly, Hubble radius) during
inflation, i.e. when $aH=k$. The CMB anisotropy measured by COBE
\cite{cobe} allows a determination of the fluctuation amplitude at the
scale, $k_{\rm COBE}^{-1}\sim\,H_0^{-1}\simeq\,3000\,h^{-1}$\,Mpc,
corresponding roughly to the size of the presently observable
universe, where $h\equiv\,H_0/100$\,km\,sec$^{-1}$\,Mpc$^{-1}$ is the
present Hubble parameter. The number of e-folds before the end of
inflation when this scale crosses the Hubble radius is
\begin{eqnarray}
 N_{\rm COBE} \equiv N_\star (k_{\rm COBE}) & \simeq 51 
  & + \ln\left(\frac{k_{\rm COBE}^{-1}}{3000h^{-1}\,{\rm Mpc}}\right) 
    + \ln\left(\frac{V_\star}{3\times10^{14}\,{\rm GeV}}\right) 
    + \ln\left(\frac{V_\star}{V_{\rm end}}\right) \nonumber \\
 && - \frac{1}{3}\ln\left(\frac{V_{\rm end}}{3\times10^{14}\,{\rm GeV}}\right) 
    + \frac{1}{3}\ln\left(\frac{T_{\rm reheat}}{10^5\,{\rm GeV}}\right),
\label{Nstar} 
\end{eqnarray}
where we have indicated the numerical values anticipated for the
various energy scales in our model. (Note that $N_{\rm COBE}$ is
smaller than the usually quoted \cite{inflrev} value of 62 because the
reheat temperature {\em must} be low enough to suppress the production
of unstable gravitinos which can disrupt primordial nucleosynthesis
\cite{success}.) The COBE observations sample CMB multipoles upto
$\sim20$, where the $l^{\rm th}$ multipole probes scales around
$k^{-1}\sim6000h^{-1}\,{\rm Mpc}/l$. The low multipoles, in particular
the quadrupole, are entirely due to the Sachs-Wolfe effect on
super-horizon scales ($k^{-1}>k_{{\rm
dec}}^{-1}\simeq\,180h^{-1}$\,Mpc) at CMB decoupling and thus a direct
measure of the primordial perturbations. However the high multipoles
are (increasingly) sensitive to the composition of the dark matter
which determines how the primordial spectrum is modified through the
growth of the perturbations on scales smaller than the horizon at the
epoch of matter-radiation equality, i.e. for $k^{-1}<k_{\rm
eq}^{-1}\simeq\,80h^{-1}$\,Mpc. Thus the normalisation of the spectrum
(\ref{deltah}) to the COBE data is sensitive to its $k$ dependence and
also on whether there is a contribution from gravitational waves to
the CMB anisotropy. The 4-year COBE data is fitted by a scale-free
spectrum, $\delta^2_{{\rm H}}\sim\,k^{n-1}$, $n=1.2\pm0.3$, with
$Q_{\rm rms}=15.3^{+3.8}_{-2.8}\,\mu$K \cite{cobe}. For a
scale-invariant ($n=1$) spectrum, $Q_{\rm rms}=18\pm1.6\,\mu$K, so
assuming that there are no gravitational waves, the amplitude for a
$\Omega=1$ CDM universe is $\delta_{{\rm H}}=1.94\pm0.14\times10^{-5}$
\cite{cobenorm}. Using eq.(\ref{deltah}), the vacuum energy at this
epoch is then given by
\begin{equation}
 V_{\rm COBE} \simeq (6.7 \times 10^{16}\,{\rm GeV})^4\ \epsilon_{\rm COBE}\ ,
\label{scale}
\end{equation}
showing that the inflationary scale is far below the Planck scale
\cite{inflrev}. A similar limit obtains, viz. $V_{\rm
COBE}\lesssim(4.9\times10^{16}\,{\rm GeV})^4$, if the observed
anisotropy is instead ascribed entirely to gravitational waves, the
amplitude of which, in ratio to the scalar perturbations, is just
\cite{inflrev}
\begin{equation}
 r = 12.4\ \epsilon\ .
\label{grav}
\end{equation}
Thus it is legitimate to study inflation in the context of an
effective field theory. We will consider the class of models in which
the evolution of the inflaton potential is dominated by a {\em single}
power at the point where the observed density fluctuations are
produced, as is the case in all of the supergravity models so far
considered. The potential then has the form
\begin{equation}
 V \sim \Lambda^4 \left[1 + c_n \left(\frac{\phi}{M}\right)^n\right]\ .
\end{equation}
In the usual model of chaotic inflation \cite{chaotic}, one has
$\phi/M\gg1$ so the first term on the rhs is negligible and $\epsilon$
and $\eta$ are small because they are proportional to
$(\phi/M)^{-2}$. The alternative possibility \cite{chaoticnew} is for
$\phi$ to have an initial vacuum expectation value (vev) much smaller
than the Planck scale during inflation, in which case the smallness of
$V'$ and $V''$, and hence $\epsilon$ and $\eta$, results from the
relative smallness of the second term on the rhs.

It is convenient to introduce a general formalism capable of
describing both cases by expanding the (slowly varying) potential
about the value $\phi^{*}$ in inflaton field space at which the
observed density fluctuations are produced. Writing
$\phi=\tilde{\phi}+\phi^{*}$ (in units of $M$) we have
\begin{equation}
 V (\phi) = \Lambda^4 \left[1 + c_1 \tilde{\phi}
                             + c_2 \tilde{\phi}^2
                             + c_3 \tilde{\phi}^3 
                             + c_4 \tilde{\phi}^4
                             + \dots \right]\ .  
\label{expand}
\end{equation}
Here we have factored out the overall scale of inflation $\Lambda$,
which we have seen must be small relative to the Planck scale $M$. The
constraints on the parameters in the potential following from the
slow-roll conditions (\ref{slowroll}) are therefore
\begin{eqnarray}
 c_1 \ll 1\ , \quad  
 c_2 \ll 1\ , \quad 
 c_3 \tilde{\phi} \ll 1\ , \quad 
 c_4 \tilde{\phi}^2 \ll 1\ , \ldots
\label{bound}
\end{eqnarray}
We now examine whether these conditions can be naturally satisfied in
supergravity theories.

\section{Natural Supergravity Inflation}

In supersymmetric theories with a single supersymmetry generator
($N=1$), complex scalar fields are the lowest components, $\phi^a$, of
chiral superfields $\Phi^a$ which contain chiral fermions, $\psi^a$,
as their other component. (We will take $\Phi^a$ to be left-handed
chiral superfields so that $\psi^a$ are left-handed fermions.) Masses
for fields will be generated by spontaneous symmetry breaking so that
the only fundamental mass scale is the Planck scale, $M$. (This is
aesthetically attractive and also what follows if the underlying
theory generating the effective low-energy supergravity theory follows
from the superstring.) The $N=1$ supergravity theory describing the
interaction of the chiral superfields is specified by the K\"{a}hler
potential
\begin{equation}
 G (\Phi ,\Phi^\dagger) = d (\Phi , \Phi^\dagger) + \ln |f(\Phi)|^2\ , 
\label{G}
\end{equation}
which yields the scalar potential
\begin{equation}
 V = {\rm e}^{d/M^2} \left[F^{A\dagger} (d_A^B)^{-1} F_B -
      3\frac{|f|^2}{M^2} \right]
 + {\rm D-terms}\ ,
\label{V}
\end{equation}
where 
\begin{equation}
 F_A \equiv \frac{\partial f}{\partial\Phi^A} 
     + \left(\frac{\partial d}{\partial\Phi^A}\right) \frac{f}{M^2}, \qquad
 \left(d_A^B\right)^{-1} \equiv 
  \left(\frac{\partial^2 d}{\partial\Phi^A \partial\Phi_B^\dagger}\right)^{-1}.
\end{equation}
Here the function $d$ sets the form of the kinetic energy terms of the
theory
\begin{equation}
 L_{\rm kin} = \frac{\partial^2 d}{\partial\phi_A \partial\phi^{\dagger B}}
  \partial_\mu \phi_A \partial^\mu \phi^{\dagger B} ,
\label{kin}
\end{equation}
while the superpotential $f$ determines the non-gauge interactions of
the theory. For canonical kinetic energy terms,
$d=\sum_A\phi_A^\dagger\phi^A$, the potential takes the relatively
simple form
\begin{equation}
 V = \exp \left({\sum_A \phi_A^\dagger \phi^A}\right)
      \left[\sum_B \left|\frac{\partial f}{\partial \phi_B}\right|^2 - 3
     \left|f\right|^2\right] .
\label{pot}
\end{equation}
In order for there to be a period of inflation, it is necessary for at
least one of the terms $|\case{\partial{f}}{\partial\phi_B}|$ to be
{\em non-zero}. However, these are precisely the order parameters for
supersymmetry so this corresponds to supersymmetry breaking during
inflation. While there are several possible mechanisms for such
breaking, it suffices for the purposes of this discussion to simply
assume that one of the terms has nonvanishing value $\Lambda^4$, where
$\Lambda$ denotes the supersymmetry breaking scale. Now expansion of
the exponential in eq.(\ref{pot}) shows that $c_2=1$ and $c_4=1$ in
eq.(\ref{expand}), in conflict with the requirements for successful
inflation (\ref{bound}). It is seen that the problem arises due to the
presence of the overall factor involving the exponential in the
potential (\ref{pot}). The same structure typically occurs even for
more general kinetic terms (see eq.\ref{V}) and it is this that has
led to the conclusion that inflation is difficult to achieve within
the context of supergravity.

In ref.\cite{success} we suggested that in theories with moduli the
problem is easily avoided. Moduli are fields in superstring theories
which, in the absence of supersymmetry breaking, have no
potential. The moduli vevs serve to determine the fundamental coulings
of the theory and for the moduli of interest here they appear in the
superpotential only in combination with non-moduli fields, serving to
determine the latter's couplings in terms of their vevs. We argued
earlier that the quadratic terms in the potential involving the
non-moduli fields such as the inflaton would be absent for special
values of these vevs and, since the resultant potential would drive
inflation, just this desired configuration would come to dominate the
final state of the universe \cite{success}. In this paper we
demonstrate that it is not even necessary to invoke such an `anthropic
principle' because there is a {\em quasi-fixed point} in the evolution
of the moduli. This ensures, for initial values in the basin of
attraction of the fixed point, that the cancellation of the quadratic
terms applies, ensuring that condition (\ref{bound}) is satisfied.

Although the moduli have a flat potential in the absence of
supersymmetry breaking, once supersymmetry is broken they may acquire
a potential through the moduli dependence of the $d$ function in the
scalar potential (\ref{V}). This is potentially disastrous for the
mechanism discussed above because such a potential would drive the
moduli vevs away from the value needed to cancel the quadratic
inflaton term. However the kinetic term often has a larger symmetry
than the full Lagrangian; for example the canonical form has an
$SU(N)$ symmetry where $N$ is the total number of chiral fields. In
this case there will be many moduli left massless even when
supersymmetry is broken because they will be (pseudo) Goldstone modes
associated with the spontaneous breaking of this
symmetry.\footnote{Another model where a Goldstone mode has been
similarly employed is `natural' inflation \cite{natural}.} These
moduli can play the role discussed above eliminating the quadratic
term in the inflaton potential.

The mechanism we propose applies to a large class of models, the only
condition being that the kinetic term does indeed have a symmetry
leading to pseudo-Goldstone modes. However it is instructive to
construct a definite model to illustrate the idea in detail. Consider
a simple case with just two moduli $\nu_{1,2}$. The canonical kinetic
term has $g_\nu=A^{\dagger}A$ where $A^{\dagger}=(\nu_1,\nu_2)$. This
clearly has an $SU(2)$ symmetry. Since moduli are not determined
before supersymmetry breaking, they may have Planck scale vevs so it
is not sufficient to keep only the lowest order (canonical)
terms. Thus we generalize the $\nu$ kinetic term, allowing for a
general functional dependence on $A^{\dagger}A$, so
$g_\nu=g_\nu(A^{\dagger}A)$, which preserves the $SU(2)$
symmetry. When supersymmetry breaking is switched on, the moduli will
develop a potential which fixes the vev $\chi$ of $A^{\dagger}A$ at a
minimum of $g$ but leaves the ratio of the vevs of $\nu_1$ to $\nu_2$
undetermined. We now introduce a chiral matter multiplet $\phi$ which
will contain the inflaton. Unlike the moduli it may have couplings in
the superpotential which can keep it in thermal equilibrium at
temperatures below the Planck scale. (For example a superpotential of
the form $f=\rho\phi^3$ will generate a term $\rho^2|\phi|^2T^2$ in
the effective potential at high temperature, which would force
$\langle\phi\rangle/M<T$.) Although such an initial condition is not
essential to our argument,\footnote{In general, the `thermal
constraint' \cite{thermal} is irrelevant to chaotic inflationary
models \cite{chaotic,chaoticnew} wherein the initial conditions are
taken to be {\em random}, as is appropriate for singlet fields.} it
does simplify the discussion, so we take the initial conditions, at a
temperature $T\approx\Lambda$, the scale of the putative inflationary
potential, to be $\langle\phi\rangle/M\ll1$,
$\langle\nu\rangle/M\sim1$. Using this we expand the $\phi$ dependence
of the kinetic function $d$ keeping only the low-order terms:
\begin{equation}
 d (\phi ,\phi^{\dagger}, \nu , \nu^{\dagger}) = 
  g_\nu (A^{\dagger}A) + \phi^{\dagger}\phi\ h (A^{\dagger}A) + 
  \kappa (\nu_1^2 \phi^2 + \nu_1^{\dagger 2} \phi^{\dagger 2}) + \ldots 
\label{ninf}
\end{equation}
where $h$ is an unknown function and $\kappa$ is a constant. The last
term on the rhs above involves only chiral or antichiral fields
separately and can be absorbed in the superpotential. In writing this
term we have assumed there is an $U(1)$ symmetry under which $\nu_1$
and $\phi$ have opposite charges while $\nu_2$ is a singlet. Thus the
larger $SU(2)$ symmetry of the kinetic term is broken by interactions,
the gauge interactions and couplings in the superpotential. The third
term on the rhs above leads, via eq.(\ref{V}), to a term in the
potential proportional to ${\rm Re}(\nu_1^2\phi^2)$. This is the first
term sensitive to the ratio of the vevs of $\nu_1$ to $\nu_2$.

Consider now the potential following from this kinetic function. One
combination of the moduli will be driven rapidly to a minimum of
$g_\nu(A^{\dagger}A)$ due to the term
$\Lambda^4g_\nu(A^{\dagger}A)$. Writing
$\nu_i=[\tilde\nu_i+\langle\nu_i\rangle]{\rm
e}^{i\theta_i/\langle\nu_i\rangle}$, the combination
$\nu=\sqrt{\tilde\nu_1^2+\tilde\nu_2^2}$ is seen to acquire a mass of
${\cal O}(\Lambda^2/M)$ through this term, while the other three
components (pseudo-Goldstone modes) remain massless. In studying the
inflationary possiblities of the potential we are only interested in
``slow'' modes, viz. those fields with masses much less than
$\Lambda^2/M$ which satisfy the inflationary constraints
(\ref{bound}). Thus we will henceforth ignore the ``fast'' mode
$\nu$. We turn now to the second and third terms of
eq.(\ref{ninf}). Setting $A^{\dagger}A$ to its vev $\chi$, we have for
the fields involving $\phi$
\begin{equation}
 d_\phi = h (\chi) |\phi|^2 \left[1 + \lambda(\tilde{\nu}_1 + 
  \langle\nu_1\rangle)^2 \cos\left(\frac{2\theta_1}{\langle\nu_1\rangle} - 
  \frac{2\theta_\phi}{\langle\phi\rangle}\right)\right]\ ,
\label{dp}
\end{equation}
where $\lambda=2\kappa/h(\chi)$ and we have expanded $\phi$ about the
point $\langle\phi\rangle$ as
$\phi=[\tilde{\phi}+\langle\phi\rangle]{\rm
e}^{i\theta_\phi/\langle\phi\rangle}$,
$\nu_i=\tilde{\nu}_i+\langle\nu\rangle$. Expanding the exponent in
eq.(\ref{V}) now yields the leading potential term for $\phi$ of the
form $V_\phi=d_\phi\Lambda^4$. The important point is that for
specific values of $\langle\nu_1\rangle$ and $\langle\nu_2\rangle$,
the term proportional to $|\phi|^2$ in eq.(\ref{dp}) may {\em vanish},
offering the possibility of a potential for $|\phi|$ which satisfies
eq.(\ref{bound}). We will expand eq.(\ref{dp}) in the neighbourhood of
this point and show that for a large range of initial values the
fields will be driven to values such that inflation does occur.

Consider first the possible slow modes of relevance to an inflationary
era. These are $|\tilde{\phi}|$ and the non-$\nu$ component of
$\tilde{\nu}_1$ and $\tilde{\nu}_2$. On the other hand the phase
$\theta\propto(\case{2\theta_1}{\langle\nu_1\rangle} -
\case{2\theta_\phi}{\langle\phi\rangle})$ is a fast field because, for
the small $\langle\phi\rangle$ of interest here, it is dominated by
the last term and has a large mass. This is readily seen by expanding
the cosine in the potential leading to the term
V$_{\theta_\phi}=\lambda\Lambda^4h(\chi)|\phi|^2\langle\nu_1\rangle^2\theta^2
\sim\lambda\Lambda^4h(\chi)\langle\nu_1\rangle^2[\case{\langle\phi\rangle\theta_1}{\langle\nu_1\rangle}-\theta_\phi]^2$,
with a piece unsuppressed by the small vev $\langle\phi\rangle$. Thus
the phase $\theta$ will flow rapidly to the minimum
$\lambda\cos\theta=-|\lambda|$. Having identified the fast and slow
variables, we now expand the potential in the latter only, about the
vev $\langle\nu_1\rangle,\langle\nu_2\rangle$ for which the quadratic
term in $d_{\phi}$ vanishes, i.e. for $\langle\nu_1\rangle,
\langle\nu_2\rangle$ satisfying
$1-|\lambda|\langle\nu_1^2\rangle=0$. This gives
$V(|\phi|,\varphi)=\Lambda^4\beta|\phi|^2\varphi$, where $\varphi$ is
the slow component of $\tilde{\nu}_1,\tilde{\nu}_2$ in the
neighbourhood of the expansion point and $\beta$ is a constant of
order unity. We wish to study the evolution of these fields for a
range of initial conditions. For this we need to know the equation of
motion, which in turn requires the form of the kinetic terms following
from eq.(\ref{kin}). For small $\langle\phi\rangle$, the dominant term
in $d$ giving the $\tilde{\phi}$ kinetic energy is just
$\tilde{\phi}^{\dagger}\tilde{\phi}$, as the $\phi$ kinetic energy
term is canonical. (Note that this applies even though the quadratic
$|\tilde{\phi}|^2$ term in the potential has been cancelled. The
reason is that the term responsible for the cancellation in the
potential is a $\phi^2$ term in $d$ and this does not affect the
kinetic term at all! This is the underlying mechanism that evades the
problems highlighted in ref.\cite{lyth}.) For small oscillations in
$\varphi$, the kinetic function can also be expanded with leading
term, $\varphi^{\dagger}\varphi$, giving canonical kinetic energy for
this field too. Thus the equations of motion for $|\tilde{\phi}|$ and
$\varphi$ are both of the canonical form.

The example presented above assumes that the kinetic term has a larger
symmetry than the full Lagrangian. However it is straightforward to
construct other examples in which the potential discussed above
follows from a symmetry of the full theory. We illustrate this in a
model with a single modulus field, $\nu$, by the choice of kinetic
function
\begin{equation}
 d (\phi, \phi^{\dagger}, \nu, \nu^{\dagger}) = 
  g_\nu (|\nu|^2) + \phi^{\dagger} \phi\ h (|\nu|^2) + 
  \kappa [\ln(\nu)\phi^2+\ln(\nu^{\dagger})\phi^{\dagger2}] + \ldots  
\label{ninf1}
\end{equation}
Here we have written the most general form of $d$ up to terms
quadratic in $\phi$, consistent with a $Z_{2}$ symmetry under which
$\phi$ is odd, and a $U(1)$ $R$-symmetry under which $\nu$ transforms
non-trivially as $\nu\rightarrow{\rm e}^{2i\zeta}\nu$ and the
superspace co-ordinates transform as $\theta\rightarrow{\rm
e}^{-i\zeta}$. In this case, the pseudo-Goldstone modes associated
with the first two terms of eq.(\ref{ninf1}) are the phases of $\nu$
and $\phi$. The latter is a fast variable but the former is a slow
variable and generates a $|\phi|^2$ term in the potential via
$\kappa\Lambda^4[\ln(\nu)\phi^2+\ln(\nu^{\dagger})\phi^{\dagger2}]\rightarrow
-|\kappa|\Lambda^4|\phi|^2[(\ln|\nu|)^2
+\case{\theta_\nu^2}{\langle|\nu|^2\rangle}]^{1/2}$. Here we have
allowed the fast variable to determine the overall sign of the term as
before. Again a choice of the phase, $\theta_\nu$, will cancel the
$|\phi|^2$ term and expanding about this point leads to the same
potential as above.

We hope these two examples have illustrated how the cancellation of
the quadratic term can occur in a wide class of models. Now we
consider the evolution of the fields for various initial
conditions. The field potential is of the form
\begin{equation}
 V (|\tilde{\phi}|, \varphi) = \Lambda^4 \left(1 +
                         \beta |\tilde{\phi}|^2 \varphi +
                         \gamma |\tilde{\phi}|^3 +
                         \delta |\tilde{\phi}|^4 + \ldots \right)
\label{vf}
\end{equation}
where we have added further terms in the expansion of $V$. The cubic
term may arise from a cubic term in the superpotential \cite{success};
this is allowed by the symmetries we have been discussing if the
additional ($U(1)$) symmetry of the $\phi$ field in an
$R$-symmetry. (Alternatively there may be another modulus with $U(1)$
charge such that a cubic term can appear in the kinetic function $d$.)
If the cubic term is not present, then the quartic term, which is
always allowed in the kinetic term by the $SU(2)$ and $U(1)$
symmetries of the model, will dominate. Note that the parameters
$\beta$, $\gamma$ and $\delta$ are all naturally of order unity. (The
parameter $\beta$ may be chosen to be positive by definition while the
parameter $\gamma$ should be negative if it is to lead to an
inflationary potential.)

We are interested in initial conditions which lead, through thermal
effects or otherwise, to $|\tilde{\phi}|$ being small but there is
nothing which constrains the initial conditions of $\varphi$. However
we note that the potential (\ref{vf}) has an infrared fixed point with
$\tilde{\phi}=\varphi=0$. Consequently, any initial value of
$\tilde{\phi}$ and $\varphi$ will be driven there if they are within
the domain of attraction, given (for positive $\beta$) by
\begin{equation}
 \varphi \geq \frac{3|\gamma|}{2\beta} \left[1 + 
  \left\{1 + \frac{4}{9}\left(\frac{\beta}{|\gamma|}\right)^2\right\}^{1/2}
  \right] |\tilde{\phi}|\ .
\end{equation}
Therefore, {\em without any fine tuning of the initial conditions}
(beyond the condition that the fields lie in this domain of
attraction), the fields are driven to fixed values and the potential
becomes a constant, driving a period of inflation. This fixed point
corresponds to a point of inflection in the potential which is
unstable with respect to small perturbations. Thus inflation is
naturally terminated by a mechanism which we believe has not been
discussed earlier. The equations of motion for $\varphi$ and
$|\tilde{\phi}|$ are
\begin{equation}
 \ddot{\varphi} + 3 H \dot{\varphi} = - \beta |\tilde{\phi}|^2, \qquad 
 |\ddot{\tilde{\phi}}| + 3H|\dot{\tilde{\phi}}| 
  = - \beta \varphi |\tilde{\phi}| + 3 |\gamma| |\tilde{\phi}|^2 ,
\label{em}
\end{equation}
so while $\varphi$ is positive, the fields are driven to the fixed
point and inflation begins. However if $\varphi$ should fluctuate and
become negative the fields will be driven away from the fixed point
thus ending inflation. (For $\beta$ negative, the reverse would be the
case.) Now the fluctuations of $\varphi$ are of order the Hawking
temperature of the De Sitter vacuum, $T_{\rm H}=H/2\pi$, thus once
$\varphi$ is driven (from the positive direction) to be of ${\cal
O}(T_{\rm H})$, fluctuations will lead to it becoming negative and end
inflation. The initial conditions for this stage are
$\varphi,|\tilde{\phi}|\sim\,H$; thereafter, as we see from
eq.(\ref{em}), $|\tilde{\phi}|$ will grow more rapidly than $\varphi$
and the cubic term in the potential will soon dominate.

We have argued that the potential of the form (\ref{vf}) arises
naturally in supergravity models with moduli such as may be expected
from the superstring. There are two distinctive features of this
potential which ensure that, after the transition to positive
$\varphi$, there will be an inflationary period yielding density
fluctuations of the magnitude observed. The first is that this
potential has a very small gradient in the neighbourhood of the origin
in field space so it generates a long period of slow-roll inflation
during which quantum fluctuations are naturally small. The second
feature is that the full potential, including higher order terms, is
governed by an overall scale, $\Lambda$. The reason is that the
potential arises from the $d$ term of eq.(\ref {G}) which, in the
absence of supersymmetry breaking, gives rise to the kinetic term and
thus does not contribute to the potential, vanishing when derivatives
are set to zero. Thus the potential is proportional to the (fourth
power of the) overall supersymmetry breaking scale, $\Lambda$. This
scale is expected to be of ${\cal O}(10^{14})$\,GeV \cite{success}
and, in conjunction with the small slope, correctly yields the
required magnitude of density fluctuations.

\section{Implications for Large-Scale Structure and CMB Anisotropy}

The inflationary period following from a potential of the form
(\ref{vf}) with no quadratic term (and $\gamma=-4$) has been closely
studied earlier \cite{success}. The field value when perturbations of
a given scale cross the Hubble radius is obtained by integrating the
equation of motion (\ref{em}) back from the end of inflation, which
occurs at $\tilde{\phi}_{\rm end}\simeq\,M/6|\gamma|$ when
$\epsilon=1$. Thus
$\tilde{\phi}_\star\simeq\,M/3{|\gamma|}[N_\star(k)+2]$ and using
eq.(\ref{Nstar}) we find a logarithmic (squared) deviation from scale
invariance for the scalar perturbations,
\begin{equation}
 \delta^2_{{\rm H}}(k) = \frac{9\gamma^2}{75\pi^2} 
  \frac{\Lambda^{4}}{M^{4}} [N_\star(k) + 2]^{4}\ .
\label{spec}
\end{equation}
This corresponds to a `tilted' spectrum, $\delta^2_{{\rm
H}}(k)\propto\,k^{n-1}$, with
\begin{equation}
 n (k) = 1 + 2\eta - 6\epsilon \simeq \frac{N_\star(k)-2}{N_\star(k)+2}\ ,
\end{equation}
i.e. $n\simeq0.92$ for $N_\star=51$ corresponding to the scales probed
by COBE \cite{success}. We emphasize that a leading cubic term in the
potential gives the ${\em maximal}$ departure from scale-invariance.
The slope of the potential is tiny,
$\epsilon=1/18\gamma^2(N_{\star}+2)^4\simeq7.0\times10^{-9}\gamma^{-2}$,
but its curvature is not: $\eta=-2/(N_{\star}+2)\simeq-0.038$.
Consequently, although the spectrum is tilted, the gravitational wave
background (\ref{grav}) is {\em negligible}. Furthermore the tilt
would be greater if $N_\star$ is smaller, for example if there is a
second epoch of `thermal inflation' when the scale-factor inflates by
$\sim20$ e-folds \cite{thermalinfl} so that the value of $N_\star$
appropriate to COBE is 31 rather than 51, and $n\simeq0.88$. We
normalize the spectrum (\ref{spec}) to the CMB anisotropy using the
expression for the (ensemble-averaged) quadrupole \cite{cmbrev},
\begin{equation}
 \frac{\langle{Q_{\rm rms}}\rangle^2}{T_0^2} = \frac{5 C_2}{4\pi} = 
   \frac{5}{4} \int_0^\infty \frac{{\rm d}k}{k}\  
   j_2^2\left(\frac{2k}{H_0}\right)\delta^2_{{\rm H}}(k)\ ,
\end{equation}
where $j_2$ is the second-order spherical Bessel function. According
to the COBE data \cite{cobe,cobenorm}, $Q_{\rm
rms}\simeq\,20\pm2\,\mu$K for $n\simeq0.9$ which fixes the
inflationary scale to be
\begin{equation}
 \frac{\Lambda}{M} \simeq 2.8 \pm 0.14 \times 10^{-4}\ |\gamma|^{-1/2}\ ,
\end{equation} 
consistent with general considerations of supersymmetry breaking {\em
during} inflation \cite{success}.

\begin{figure}[htb]
\epsfxsize5.5in\epsffile{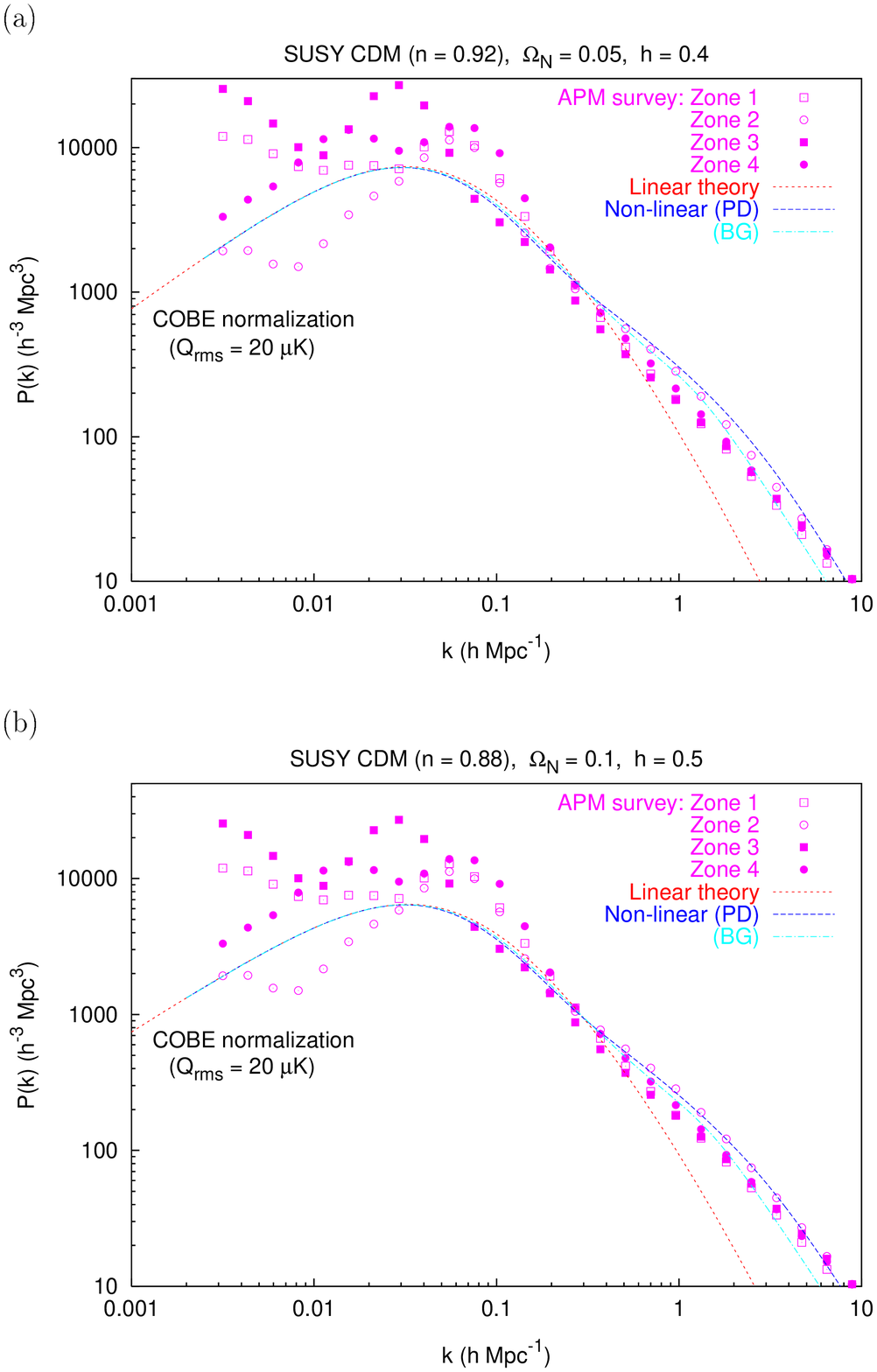}
\caption{Predicted power spectrum of density fluctuations in cold dark
 matter compared with data from the APM survey. The dotted line shows
 the linear spectrum, and the dashed lines the non-linear evolution
 according to two different prescriptions. The spectra are normalized
 to COBE adopting (a) $N_{\rm COBE}=51,\ \Omega_{\rm N}=0.05,\ h=0.4$,
 and (b) $N_{\rm COBE}=31,\ \Omega_{\rm N}=0.1,\ h=0.5$.}
\label{ps}
\end{figure}

The spectrum of the (dimensionless) rms mass fluctuations after matter
domination (per unit logarithmic interval of $k$) is given by
\cite{structure}
\begin{equation}
 \Delta^2 (k) \equiv \frac{k^3 P(k)}{2\pi^2} 
              = \delta^2_{{\rm H}}(k)\ T^2 (k) \left(\frac{k}{aH}\right)^4\ ,
\end{equation}
where $P(k)$ is the usual power spectrum and the `transfer function'
$T(k)$ takes into account that linear perturbations grow at different
rates depending on the relation between their wavelengths, the Jeans
length and the Hubble radius. For CDM we use the parametrization
\cite{structure},
\begin{equation}
 T (k) = \left[1 + \left\{a k + (b k)^{3/2} + (c k)^2
          \right\}^{\nu}\right]^{-1/\nu}
\end{equation}
with $a=6.4\Gamma^{-1}h^{-1}{\rm Mpc}$, $b=3\Gamma^{-1}h^{-1}{\rm
Mpc}$, $c=1.7\Gamma^{-1}h^{-1}{\rm Mpc}$ and $\nu=1.13$, where the
`shape parameter' is $\Gamma\simeq\Omega{h}\,{\rm e}^{-2\Omega_{N}}$
\cite{peadodd}. For `standard' CDM, $h=0.5$ and $\Omega_{N}=0.05$
\cite{structure}. However, observational uncertainties still permit
the Hubble parameter to be as low as 0.4 \cite{hrev} and the nucleon
density parameter $\Omega_{\rm N}$ may be as high as
$\sim0.033h^{-2}$, taking into account the recent upward revision of
the $^4$He mass fraction \cite{bbn}. We show $P(k)$ for
$\Omega_{N}=0.05,\,0.1$ and $h=0.4,\,0.5$ in figure~\ref{ps}, having
taken account of non-linear gravitational effects at small scales
using the prescriptions of ref.\cite{peadodd2} (PD) and
ref.\cite{nonlin} (BG). The tilt in the primordial spectrum which {\em
increases} logarithmically with decreasing scales allows a good fit to
the data points obtained \cite{apm} from the angular correlation
function of APM galaxies, if the Hubble parameter (nucleon density)
are taken to be at the lower (upper) end of the allowed range. (We
have not allowed for the evolution of clustering in the APM data which
is estimated to systematically raise the data points by $14\%$ for an
unbiased $\Omega=1$ CDM universe \cite{enrique}.) We have shown the
data separately for the 4 zones of the APM survey to illustrate that
there are large errors \cite{apm} for $k\lesssim0.1h\,{\rm Mpc}^{-1}$;
thus the apparent discrepancy here requires further investigation
\cite{progress}. However at small scales, the data sets agree well and
reveal the expected characteristic ``shoulder'' due to non-linear
evolution which is reproduced by our model.  Other studies of tilted
spectra \cite{tilt,liddle} focussed on the linear evolution and/or
used a compendium \cite{peadodd} of data from different surveys
(having different systematic biases) rather than one set of high
quality data. We conclude that the problem with the excess power on
small scales in the COBE-normalized standard CDM model \cite{white} is
naturally alleviated in supergravity inflation, as anticipated earlier
\cite{success,eps}, with {\em no need for a component of hot dark
matter}.

\begin{figure}[htb]
\epsfxsize5.5in\epsffile{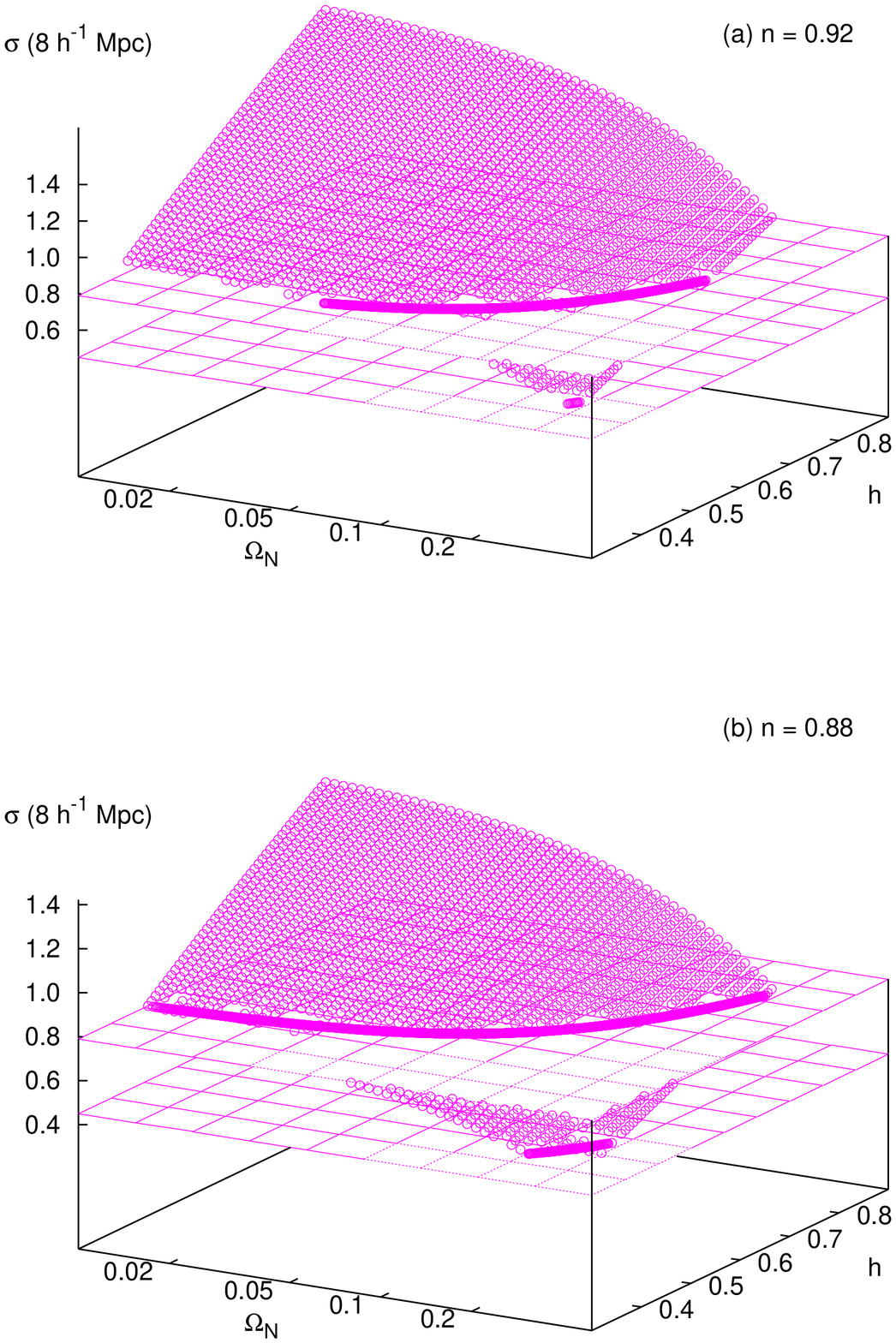}
\caption{Predicted value of the variance of the density field smoothed
 over a sphere of radius $8\,h^{-1}$\,Mpc for (a) $N_{\rm COBE}=51$
 and (b) $N_{\rm COBE}=31$, as a function of the Hubble parameter and
 the nucleon density parameter. The region within the marked contours
 is consistent with the observational limits (horizontal planes) inferred
 from rich clusters of galaxies.}
\label{sigma8}
\end{figure}

We also quote some averaged quantities of observational interest for
this model. A common measure of large-scale clustering is the
variance, $\sigma(R)$, of the density field smoothed over a sphere of
radius $R$, usually taken to be $8\,h^{-1}$\,Mpc, given in terms of
the matter density spectrum by
\begin{equation}
 \sigma^2 (R) = \frac{1}{H_0^4} \int^\infty_0 W^2 (kR)\ 
                 \delta^2_{\rm H}(k)\ T^2(k)\ k^3\ {\rm d} k\ , 
\end{equation}
where a `top hat' smoothing function,
$W(kR)=3\left[\frac{\sin(kR)}{(kR)^3}-\frac{\cos(kR)}{(kR)^2}\right]$,
has been used. As seen from figure~\ref{sigma8}, the observational
value of $\sigma\,(8 h^{-1}{\rm Mpc})=0.60^{+0.19}_{-0.15}$ ($95\%$
c.l.), inferred from the abundances of rich clusters of galaxies
\cite{white,viana} favours high tilt, high $\Omega_{\rm N}$ and low
$h$. For the two models shown in figure~\ref{ps} we find,
\begin{eqnarray}
 \sigma (8 h^{-1}{\rm Mpc}) &= 0.78 &\pm 0.08\ 
  (N_{\rm COBE}=51,\ \Omega_{\rm N}=0.05,\ h=0.4)\ , \nonumber \\ 
                            &= 0.75 &\pm 0.08\ 
  (N_{\rm COBE}=31,\ \Omega_{\rm N}=0.1,\ h=0.5)\ .
\end{eqnarray}
Another interesting quantity is the smoothed peculiar velocity field
or `bulk flow',
\begin{equation}
 \sigma_v^2(R) = \frac{1}{H_0^2} \int^\infty_0 W^2(kR)
                 \ {\rm e}^{-(12\,h^{-1} k)^2}\ 
                 \delta^2_{\rm H}(k)\ T^2(k)\ k\ {\rm d}k\ ,
\end{equation}
where, for direct comparison with observations, we have applied an
additional gaussian smoothing on $12 h^{-1}{\rm Mpc}$. With the same
parameters as above,
\begin{eqnarray}
 \sigma_v (40 h^{-1}{\rm Mpc}) &= 383 &\pm 38\ {\rm km\ sec}^{-1}\  
  (N_{\rm COBE}=51,\ \Omega_{\rm N}=0.05,\ h=0.4), \nonumber \\ 
                               &= 320 &\pm 32\ {\rm km\ sec}^{-1}\ 
  (N_{\rm COBE}=31,\ \Omega_{\rm N}=0.1,\ h=0.5).
\end{eqnarray}
to be compared with the POTENT III measurement of
$\sigma_v(40h^{-1}{\rm Mpc})=373\pm50$\,km~sec$^{-1}$
\cite{potent}. We do not consider constraints coming from the
abundances of collapsed objects at high redshift such as
Lyman-$\alpha$ clouds and quasars \cite{liddle,repeat}, as this
involves many astrophysical uncertainties.
 
\begin{figure}[htb]
\epsfxsize5.5in\epsffile{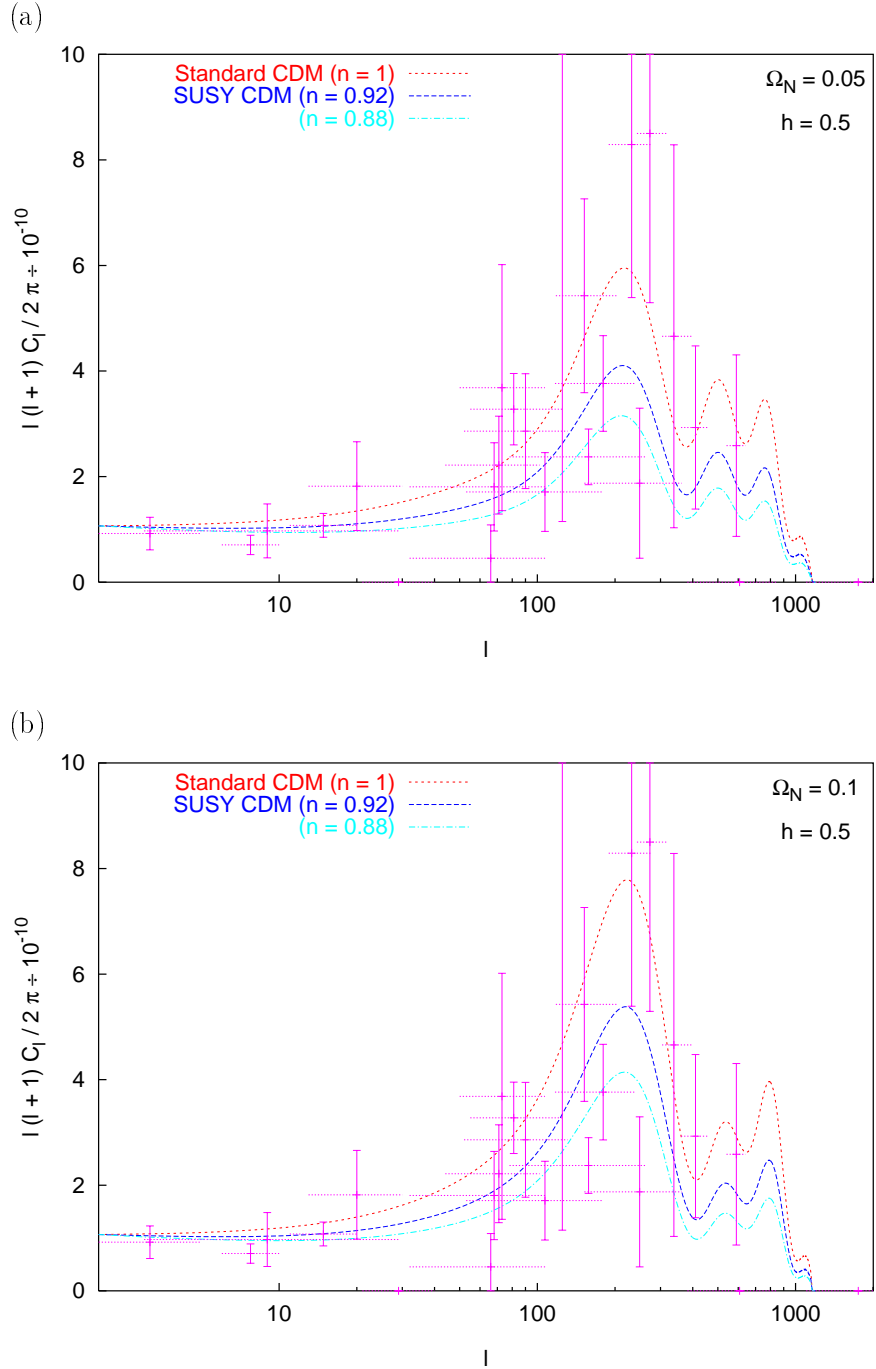}
\caption{Predicted angular power spectrum of CMB anisotropy normalized
 to COBE and compared with current data, adopting (a) $\Omega_{\rm
 N}=0.05$, and (b) $\Omega_{\rm N}=0.1$, both with $h=0.5$. The
 standard scale-invariant spectrum (full line) is compared with the
 tilted spectra from supergravity inflation for $N_{\rm COBE}=51$
 (dashed line) and $N_{\rm COBE}=31$ (dotted line).}
\label{cmb}
\end{figure}

An unambiguous test of the model is the predicted CMB anisotropy. To
compute this accurately requires numerical solution of the coupled
linearized Boltzmann, Einstein and fluid equations for the
perturbation in the photon phase space distribution. We use the
COSMICS computer code \cite{bert} developed to calculate the angular
power spectrum using the primordial scalar fluctuation spectrum
(\ref{spec}). These programmes include a careful treatment of the
hydrogen recombination and the decoupling of the matter and radiation,
a full treatment of Thompson scattering, and a full computation of all
relativistic shear stresses of photons and neutrinos. The first 1000
multipoles are plotted in figure~\ref{cmb}, taking $\Omega_{\rm
N}=0.05,\ 0.1$, along with a compendium of recent observational data
\cite{cmbrev}, and the prediction of standard CDM is shown for
comparison. The height of the first `Doppler peak' is preferentially
boosted for the higher value of $\Omega_{\rm N}$ and this is favoured
by the CMB observations in conjunction with the large-scale structure
data, as has been noted independently \cite{repeat}. For a given value
of $\Omega_{\rm N}$ the effect of the spectral tilt is to suppress the
heights of all Doppler peaks. Although present ground-based
observations are inconclusive, this prediction will be definitively
tested by the forthcoming satellite-borne experiments, MAP and
COBRAS/SAMBA.

{\bf Acknowledgements:} 
We are grateful to George Efstathiou and especially to Enrique
Gaztan\~{a}ga, for providing the APM data, and for many stimulating
discussions. We thank David Lyth for motivating us to clarify and
extend our previous work and Ed Copeland for helpful comments. This
research was supported by the EC Theoretical Astroparticle Network
CHRX-CT93-0120 (DG12 COMA).

\end{document}